\documentclass[twocolumn,showpacs]{revtex4}

\def\v#1{\mathbf{#1}}

\usepackage{graphics,epsfig,graphicx}
\usepackage{color}
\usepackage{latexsym}

\begin{document}

\title{A predicted ``Faraday oscillation'' in photoexcited semiconductors}

\author{M. Combescot, O. Betbeder-Matibet  \\
\it{ Institut des NanoSciences de Paris,}\\
\it{ Universit\'{e} Pierre et Marie Curie-Paris 6,CNRS, UMR 7588,}\\
\it{Campus Boucicaut, 140 rue de Lourmel,75015 Paris}}

\begin{abstract}
While, for semiconductors photoexcited by a circularly polarized pump, the
polarization plane of a linearly polarized probe has been shown to
rotate, we here predict a spectacular change when the pump beam is
linearly polarized, from Faraday rotation to
``\emph{Faraday oscillation}'', the oscillation of the polarization plane
going along a change of the photon polarization from linear to
elliptical. This effect, which reduces to zero when the probe field
is parallel or perpendicular to the pump field, comes from
\emph{coherence} between  the real excitons created by the pump
and the virtual exciton coupled to the unabsorbed probe --- as easy to
see from the Shiva diagrams which represent the many-body physics taking
place in this coupled photon-composite-exciton system.
\end{abstract}
\maketitle

Faraday rotation is a physical effect known for quite a long time [1]. It
takes place in ``optically active media'' which have different refractive
indices
$n_{\pm}$ for photons with circular polarization:
the two parts of a linearly polarized probe beam then
travel at different speeds, making the polarization plane of this probe
beam rotate around the propagation axis, with a rotation proportional
to
$(n_+-n_-)$. The refractive index difference physically comes from a
dissymetry of the material which can preexist or be induced. The standard
way to induce this dissymetry is through a magnetic field which produces
an energy difference between electrons with up and down spins. In
semiconductors, this dissymetry can also be produced by the absorption of
a circularly polarized pump beam [2-13]. In a recent work [14], we have
shown, by using the many-body theory we have developed for composite
bosons [15], that the real excitons created by the pump interact with the
virtual excitons coupled to the
$\sigma_{\pm}$ parts of the unabsorbed probe, not only through Coulomb
interaction, but mainly through the Pauli exclusion principle between the
fermionic components of these excitons. With a
$\sigma_+$ pump acting on quantum well, this exclusion --- which is
directly linked to the exciton composite nature --- only exists with the
virtual excitons coupled to the $\sigma_+$ part of the linearly polarized
probe. The resulting difference in the response function $S_{\pm}$ to
$\sigma_{\pm}$ photons leads to $n_+\neq n_-$.

We here consider a far more subtle situation: If instead of using
$\sigma_+$ photons, we create excitons through the absorption of a pump
beam having a linear polarization along $\v x$, a na\"{\i}ve thinking
would lead to say that the $\sigma_+$ and $\sigma_-$ parts of a linearly
polarized probe see the $\v x$ excitons --- which are half $\sigma_+$
and half $\sigma_-$ --- in the same way, so that the refractive indices
$n_{\pm}$ should be the same. This na\"{\i}ve thinking however forgets the
importance of coherence between the $\sigma_{\pm}$ parts of the pump and
the probe: It is somewhat obvious that, in the same way as excitons
created by a $\sigma_+$ pump beam act differently on the
$(\sigma_+,\sigma_-)$ parts of a linearly polarized probe, excitons
created by a $\v x$ pump must affect the $(\v x,\v y)$ parts of the light
differently.

\begin{figure}[t]
\centerline{\scalebox{0.8}{\includegraphics{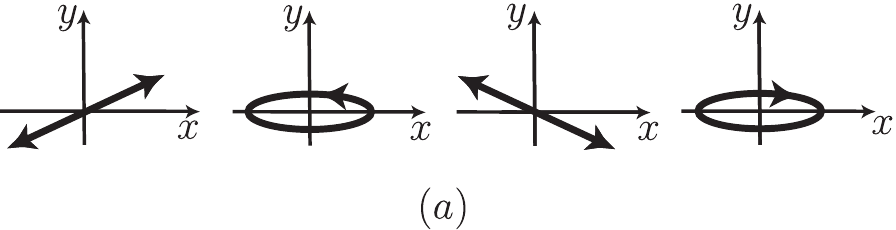}}}
\centerline{\scalebox{0.8}{\includegraphics{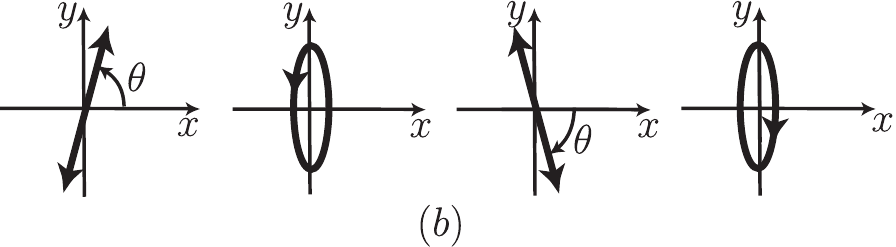}}}
\centerline{\scalebox{0.8}{\includegraphics{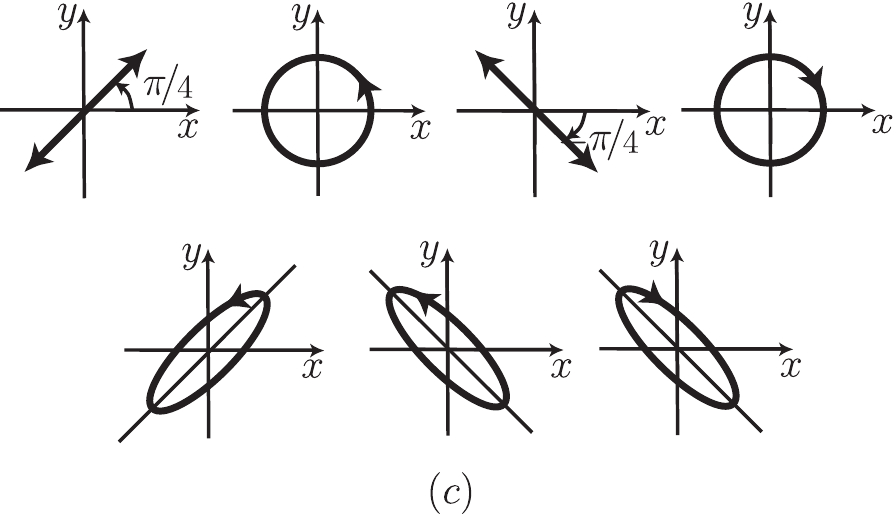}}}
\caption{(a) Time evolution of the probe photon ellipticity when, for
$t=0$, the probe has a linear polarization along $\theta_0$ with
$0<\theta_0<\pi/4$, while the pump excitons are polarized along $\v
x$. The ellipticity change goes along an oscillation of the polarization
main axis around $\v x$. (b) Same as (a) for $\pi/4<\theta_0<\pi/2$, the
oscillation taking place around $\v y$. (c) When $\theta_0=\pi/4$, the
probe photon polarization goes from linear to circular,
its main axis staying at $\pi/4$ from $(\v x,\v y)$. It then returns to
linear but with a main axis at $(-\pi/4)$. And so on \ldots}
\end{figure}

In this letter, we predict a spectacular change when the polarization of
the pump goes from circular to linear. Indeed, a rotation of the
polarization plane implies to break
the clockwise-counterclockwise invariance of the system. This happens
with excitons created by $\sigma_+$ photons, but not by
a linearly polarized pump.
With such a pump, the probe polarization plane has no reason to rotate
one way or the other; if something happens, this cannot be a rotation. We
here show that, in the case of a
$\v x$ pump, an unabsorbed probe beam polarized along $\v x$ or $\v y$
stays unchanged, while for any other direction $\theta_0\neq (0,\pi/2)$,
the polarization changes from linear to elliptical,
the ellipse main axis oscillating around $\v x$ or $\v
y$: When
$0<\theta_0<\pi/4$, the ellipticity of the probe photon first increases
while the main axis rotates from $\theta_0$ to 0. This ellipticity then
decreases while the main axis keeps rotating from 0 to $-\theta_0$ where
it turns linear again (see Fig.1(a)). For $\pi/4<\theta_0<\pi/2$, a
similar oscillation takes place around $\v y$ (see Fig.1(b)). Continuity
between these two oscillations is obtained for $\theta_0=\pi/4$ with a
probe beam which goes from linear to circular, its main axis staying along
$\theta_0=\pi/4$. The probe photon ellipticity then shrinks to become
linear again, but with a main axis along
$(-\pi/4)$, where it returns to circular, the main
axis staying along $(-\pi/4)$, and so on\ldots (see Fig.1(c)).

\noindent\textbf{Physical idea.}
This behavior follows from the idea that eigenstates for
photons coupled to excitons linearly polarized along $\v x$, have an $\v
x$ or $\v y$ polarization. The time evolution of a linear polarization
along $\theta_0$ for $t=0$ then reads
\begin{equation}
\v e_t=\v x\,e^{-i\Omega_xt}\cos\theta_0+\v
y\,e^{-i\Omega_yt}\sin\theta_0\ .
\end{equation}
By writing this polarization as
\begin{equation}
\v e_t=e^{i\varphi_t}(\v X_t\,\cos\xi_t+ i\v Y_t\,\sin\xi_t)\ ,
\end{equation}
which describes photons with ellipticity $\xi_t$ and main axis $(\v X_t,\v
Y_t)$, we find from $(\v e_t.\v X_t)$ and $(\v e_t.\v Y_t)$,
\begin{equation}
e^{2i\xi_t}=\cos 2\theta_0\cos 2\alpha_t+\sin 2\theta_0(\sin
2\alpha_t\cos\Delta_t+i\sin\Delta_t)\ ,
\end{equation}
where $\Delta_t=(\Omega_x-\Omega_y)t$ while $\alpha_t$ is such that $\v
X_t=\v x\,\cos\alpha_t+
\v y\,\sin\alpha_t$.

Equation (3) readily gives the time evolution of the
ellipticity as
\begin{equation}
\sin 2\xi_t=\sin 2\theta_0\sin\Delta_t\ .
\end{equation}
This shows that photons stay linearly polarized for $\theta_0=(0,\pi/2)$,
\emph{i.e.}, when the polarization is along $\v x$ or $\v y$, while for
$\theta_0=\pi/4$, the photon ellipticity increases up to circular, which
is reached for $\Delta_t=(\pi/2, 3\pi/2,\cdots)$.

From the modulus of eq.\ (3), we find the time evolution of the photon
polarization main axis as
\begin{equation}
\tan 2\alpha_t=\tan 2\theta_0\cos\Delta_t\ .
\end{equation}
This shows that $\v X_t$ oscillates from $\alpha_t=\theta_0$ to
$\alpha_t=-\theta_0$ when $0<\theta_0<\pi/4$, while for
$\pi/4<\theta_0<\pi/2$, it oscillates from $\theta_0$ to $(\pi-\theta_0)$
(in order to stay unchanged for $\theta_0=\pi/2$). The intermediate value
$\theta_0=\pi/4$ can appear as singular, since $\alpha_t$ jumps from
$\pi/4$ to $-\pi/4$ when
$\Delta_t$ passes $\pi/2$. However, for $\Delta_t=\pi/2$,
the photon polarization is then circular, so that the ``main axis''
$\pi/4$ and $-\pi/4$ are totally equivalent. Equations (4,5) thus
support the time evolution of the probe polarization shown in Fig.1

\begin{figure}[t]
\vspace{-1cm}
\hspace{-0.5cm}
\centerline{\scalebox{0.36}{\includegraphics{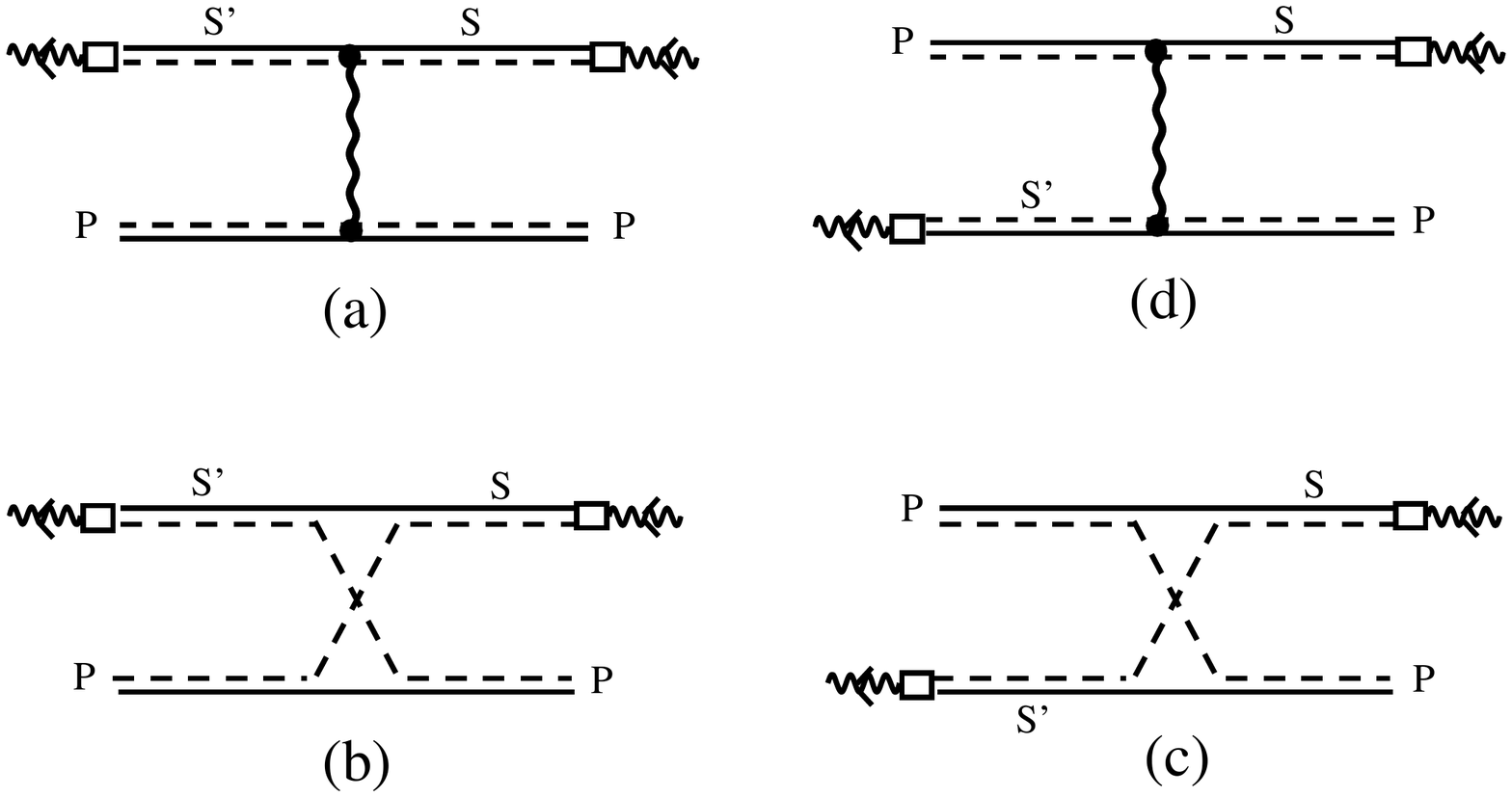}}}
\vspace{-2.5cm}
\caption{Possible processes between $\sigma_S$ photons (wavy lines)
coupled to $(S=\pm 1)$ virtual excitons, these composite excitons
interacting with one pump exciton $P$ having a polarization
$\cos\xi'_p(+1)+\sin\xi'_p(-1)$ by Coulomb interaction (diagrams a,d)
or just by carrier exchange (diagrams b,c).}
\end{figure}

Before going further, let us mention that the standard Faraday rotation,
obtained with circularly polarized pump photons, can be recovered
along the same line. Indeed, the $t=0$ probe photon polarization
given in eq.\ (1), reads in terms of the $\sigma_{\pm}$
polarization
$\v e_{\pm}=(\v x\pm i\v y)/\sqrt{2}$ as
\begin{equation}
\v e_{t=0}=(e^{-i\theta_0}\,\v e_+ +e^{i\theta_0}\,\v e_-)/\sqrt{2}\ .
\end{equation}
When the photon eigenstates have a circular polarization --- as obtained
with excitons created by a $\sigma_+$ pump  ---  the
probe photon polarization evolves according to
\begin{equation}
\v e_t=\left[e^{-i\theta_0}\v e_+\,e^{-i\Omega_+t}+e^{i\theta_0}\v
e_-\,e^{-i\Omega_-t}
\right]/\sqrt{2}\ .
\end{equation}
Within an irrelevant phase factor $\varphi_t=-(\Omega_++\Omega_-)t/2$,
this corresponds to probe photons staying linearly polarized with a
polarization plane rotating as $\theta_t=\theta_0+(\Omega_+-\Omega_-)t/2$.

\noindent\textbf{Photon eigenstates.}
Let us now show the link between the photon eigenstates and the pump
polarization $\v e_p$. As $\v e_p=\v x\,\cos\xi_p+i\v y\,\sin\xi_p$
also reads $\v e_+\,\cos\xi'_p+\v e_-\,\sin\xi'_p$, with
$\xi'_p=\pi/4-\xi_p$, this pump gives rise, if we
neglect spin relaxation, to $N$ coherent excitons
$B_P^\dag=\cos\xi'_pB_{o,+1}^\dag+\sin\xi'_pB_{o,-1}^\dag$, where
$B_{o,S}^\dag$ creates a ground state exciton with spin $S$. An
unabsorbed probe photon with polarization
$S=\pm 1$ interacts with these pump excitons through the virtual excitons
$B_{i,S}^\dag$ to which they are coupled. The lowest order term in the
pump exciton density comes from processes in which one virtual exciton
interacts with
\emph{one} among
$N$ pump exciton. As shown in our previous work on Faraday rotation [14],
these processes are represented by the four Shiva diagrams [16] of Fig.2,
in which the excitons coupled to the ``in'' and ``out'' probe photons have
2, 1 or 0 common carriers --- with possibly some additional Coulomb
processes.

Due to spin conservation when the ``in'' and ``out''
excitons are made from the same pairs, we readily see that the diagram
(2a) has a polarization prefactor\linebreak
$\delta_{S',S}(\cos^2\xi'_p+\sin^2\xi'_p)$ while the prefactor for the
diagram (2d) is
\begin{equation}
\left(\cos\xi'_p\delta_{S',+1}+\sin\xi'_p\delta_{S',-1}\right)
\left(\cos\xi'_p\delta_{+1,S}+\sin\xi'_p\delta_{-1,S}\right)\ .
\end{equation}
For the diagrams (2b,2c), in which the excitons exchange
one carrier, we must remember that as, in quantum wells, carrier exchanges
between $(+1,-1)$ excitons lead to dark states $(\pm 2)$,
exchanges can only exist between the $(+1)$ parts or the $(-1)$
parts of these excitons [17]. This leads to the same
polarization prefactor $\delta_{S',S}(\delta_{S,+1}\cos^2\xi'_p+
\delta_{S,-1}\sin^2\xi'_p)$ for the diagrams (2b,2c).

By calling $(D_a,D_d)$ the orbital parts of the direct diagrams (2a,2d)
and
$(X_b,X_c)$ the orbital parts of the exchange diagrams (2b,2c), the
eigenstates for probe photons coupled to
pump excitons $P=\cos\xi'_p(+1)+\sin\xi'_p(-1)$, are obtained in the
$(+1,-1)$ basis, from
\begin{equation}
\left|\begin{array}{ll}D_a+X\cos^2\xi'_p-\Omega\ \ &D_d\sin\xi'_p
\cos\xi'_p\\D_d\sin\xi'_p\cos\xi'_p&D_a+X\sin^2\xi'_p-\Omega 
\end{array}\right|=0\ ,
\end{equation}
where $X=X_b+X_c+D_d$, the only contribution not in $\delta_{S',S}$ coming
from diagram (2d) through eq.\ (8). Equation (9) shows that (i) for a
$\sigma_+$ pump,
$\xi_p=\pi/4$,
\emph{i.e.},
$\xi'_p=0$, the photon eigenstates are $S=\pm 1$:
They have a circular polarization, with an energy splitting
$\Omega_+-\Omega_-=X$. (ii) For a pump linearly polarized along $\v x$,
$\xi_p=0$,
\emph{i.e.}, $\xi'_p=\pi/4$, the
eigenstates
$[(+1)\pm (-1)]/\sqrt{2}$ have linear polarizations along $\v x$
and $\v y$, the energy splitting being $\Omega_x-\Omega_y=D_d$.

Consequently, the
physical processes responsible for the Faraday rotation induced by a
$\sigma_+$ pump, which makes $\Omega_+\neq \Omega_-$, are the ones in
which the pump exciton and the virtual exciton coupled to the probe
exchange 1 or 2 carriers (see Figs.2(b,c,d)), as previously found using
another procedure [14]. On the opposite, the Faraday oscillation produced
by an $\v x$ pump which makes $\Omega_x\neq \Omega_y$, only comes from the
process of Fig.2(d) in which \emph{the two excitons exchange their two
carriers}.

\begin{figure}[t]
\vspace{-0.5cm}
\centerline{\scalebox{0.5}{\includegraphics{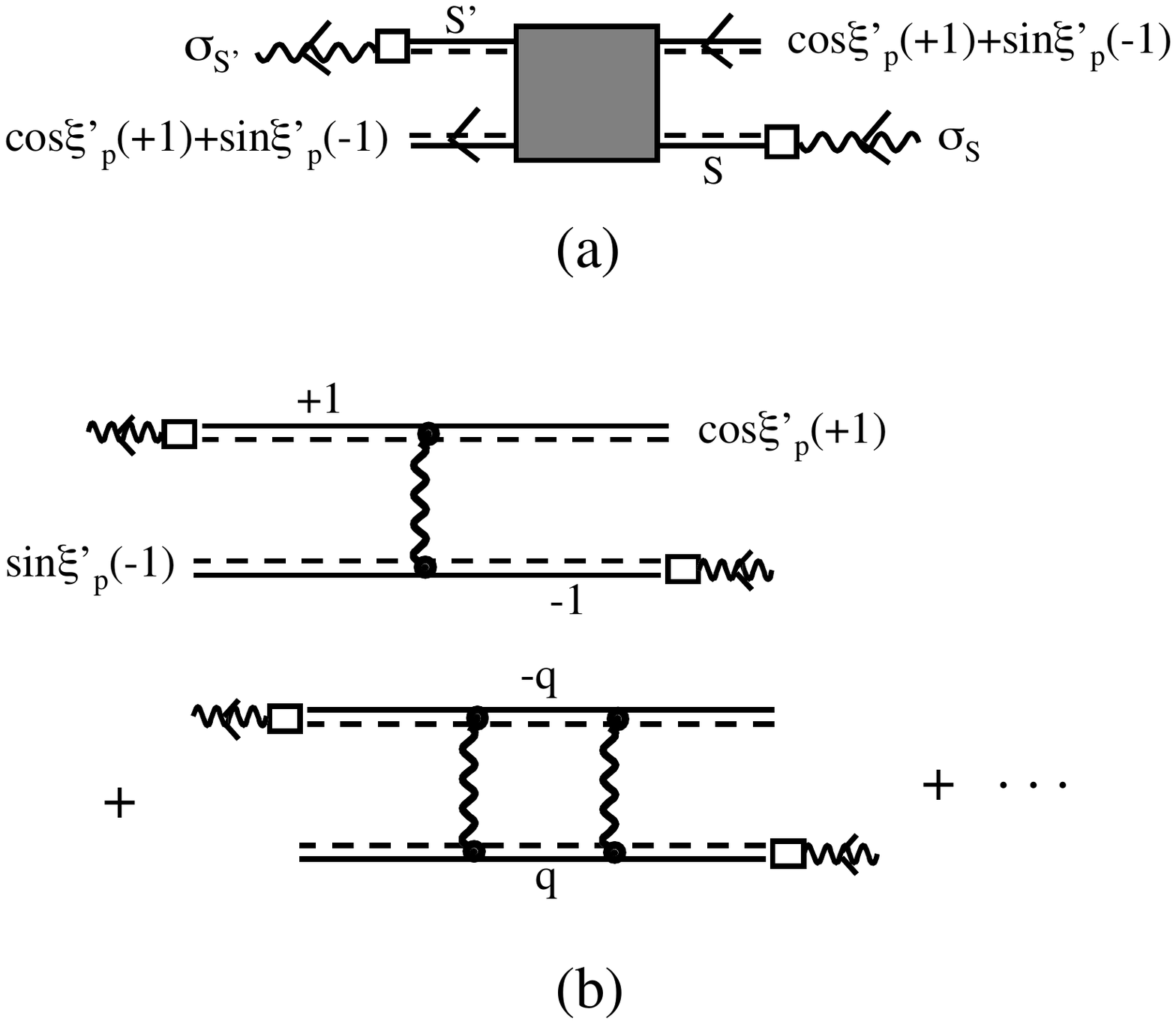}}}
\vspace{-5.5cm}
\caption{(a) Shiva diagram for the coupling $\Delta_{S',S}^{(1)}$ of one
pump exciton having a polarization
$\cos\xi'_p(+1)+\sin\xi'_p(-1)$ with one
unabsorbed photon $\sigma_S$ coupled to a virtual exciton with spin $S$.
(b) Direct Coulomb process, with one or two Coulomb interactions,
responsible for the non-diagonal coupling
$\Delta_{+1,-1}^{(1)}$ producing the Faraday oscillation.}
\end{figure}

\noindent\textbf{Formalism.}
Let us now outline the formalism which leads to eq.\ (9) for the
eigenstates of a probe photon in the presence of $N$ coherent excitons
$B_P^\dag$. In the absence of semiconductor-photon coupling, the two
states $|\phi_S\rangle=\alpha_S^\dag|v\rangle \otimes B_P^{\dag
N}|v\rangle$, where $\alpha_S^\dag$ creates a photon with polarization
$S=\pm 1$, are degenerate in energy,
$(H_0-\mathcal{E}_0)|\phi_S\rangle=0$. The eigenstates of the coupled
system $(H-\mathcal{E})|\psi\rangle=0$, with
$H=H_0+W$,
$W=U+U^\dag$ and $U^\dag=\sum_{j,S}\mu_jB_{j,S}^\dag\alpha_S$, are easy to
obtain by writing $|\psi\rangle$ as
\begin{equation}
|\psi\rangle=\left(|\tilde{\phi}_{+1}\rangle\langle\tilde{\phi}_{+1}|+
|\tilde{\phi}_{-1}\rangle\langle\tilde{\phi}_{-1}|+P_{\perp}\right)|\psi
\rangle\ ,
\end{equation}
where $|\tilde{\phi}_S\rangle$ is the normalized state
$|\phi_S\rangle$, while $P_{\perp}$ is the projector over the subspace
perpendicular to $|\tilde{\phi}_{\pm 1}\rangle$. If we insert this
$|\psi\rangle$ into the Schr\"{o}dinger equation and we multiply it by
$P_{\perp}$, we get, to lowest order in the semiconductor-photon coupling,
\begin{equation}
P_{\perp}|\psi\rangle\simeq
P_{\perp}\,\frac{1}{\mathcal{E}_0-H_0}\,P_{\perp}W\sum_S
|\tilde{\phi}_S\rangle\langle\tilde{\phi}_S|\psi\rangle\ .
\end{equation}
If we now use this $P_{\perp}|\psi\rangle$ into eq.\ (10) and multiply
the Schr\"{o}dinger equation by $\langle\tilde{\phi}_{\pm 1}|$, we get
two equations for $\langle\tilde{\phi}_{\pm 1}|\psi\rangle$. They have a
nonzero solution for
\begin{equation}
\left|\begin{array}{ll}\mathcal{E}_0+\Delta_{+1,+1}-\mathcal{E}\
\ \ &\Delta_{1,-1}\\\Delta_{-1,+1} &\mathcal{E}_0+\Delta_{-1,-1}
-\mathcal{E}\end{array}\right|=0\ ,
\end{equation}
where $\Delta_{S',S}$ is given by
\begin{equation}
\Delta_{S',S}=\langle\tilde{\phi}_{S'}|U\,\frac{1}{\mathcal{E}_0-H_0}\,U^\dag|
\tilde{\phi}_S\rangle\ .
\end{equation}

The state $U^\dag|\tilde{\phi}_S\rangle$ contains $N$ pump excitons and
one virtual exciton with spin $S$ coupled to a $S$ photon.
The contribution to $\Delta_{S',S}$ linear in the exciton density,
corresponds to the interactions of this virtual exciton with one among
$N$ pump excitons [14]. For photon detuning close enough to the exciton
ground state, in order to only keep the photon coupling to this exciton,
$\Delta_{S',S}$ reduces to
\begin{equation}
\Delta_{S',S}^{(1)}= N|\mu_o|^2\langle
v|B_PB_{oS'}\,\frac{1}{\delta+2E_o-H_{sc}}\,B_{oS}^\dag
B_P^\dag|v\rangle\ ,
\end{equation}
where $E_o$ is the exciton ground state energy, $\delta$ the photon
detuning and $H_{sc}$ the semiconductor Hamiltonian. $\Delta_{S',S}^{(1)}$
corresponds to the diagram of Fig.3(a).
Since the two-pair-eigenstate spectrum is not known,
$\Delta_{S',S}^{(1)}$ cannot be calculated exactly for any detuning.
Eq.(14) already shows that the smaller the detuning, the larger
the coupling. We can also say that, as carrier exchanges between
excitons
$(+1)$ and
$(-1)$ lead to dark states [17], the non-diagonal coupling
$\Delta_{1,-1}^{(1)}$ responsible for the Faraday oscillation can
only come from direct Coulomb processes in which the excitons $B_{o,1}$
and
$B_{o,-1}^\dag$ are made with different electron-hole pairs, like in
diagram (2d). However as, for exciton momenta close to zero, the direct
Coulomb scattering
$\xi\left(^{o\ o}_{o\ o}\right)$, which corresponds to the
first diagram of Fig.3(b), reduces to zero [18],
$\Delta_{1,-1}^{(1)}$ at large detuning is controlled by direct
processes in which enter \emph{two} Coulomb interactions at least (see
Fig.3(b)). Dimensional
arguments then lead to a non-diagonal
coupling
$\Delta_{1,-1}^{(1)}$ in $(R_X/\delta)^2$ compared to the diagonal
coupling
$\Delta_{1,1}^{(1)}$, controlled by exchange processes in the absence
of Coulomb interaction, \emph{i.e.}, diagrams (2b,2c). Consequently, at
large detuning, the energy splitting producing the Faraday oscillation is
much smaller than the one responsible for the Faraday rotation:
To observe this new Faraday oscillation, we thus need very good
samples in order to have a narrow exciton line to possibly approach
the exciton resonance without sizeable residual absorption.

Experiments on Faraday rotation or Kerr effect in photoexcited
semiconductors have been reported by various groups [2,3,5-13,19,20]].
In these experiments, the probe photons  are at resonance, while for a
``pure'' effect, we should use unabsorbed photons in order to be coupled
to virtual excitons. These experiments however show that, with a $\v x$
pump, photons polarized along $\v x$ or $\v y$ stay unchanged [19,20], in
agreement with the present work.

Note that, while Faraday rotation here appears through
the time evolution of the probe polarization in the presence of pump
excitons, this effect is usually measured when passing through a
photoexcited sample. The two points of view can be related through the
time
$t$ the light travels in the sample. Also note that the present approach
allows to point out the role of coherence between pump and probe
excitons, in contrast to  standard approaches in which the response
functions to
$\sigma_+$ and
$\sigma_-$ photons are calculated \emph{separately}.

\noindent\textbf{Conclusion.} The many-body theory for
composite bosons we have recently constructed [15] is quite appropriate to
predict subtle polarization effects like the ones induced by exciton
coherence. While the precise understanding of the last part of this
letter requires some background on this theory --- which can be found in
various previous works [15] --- the spectacular change from Faraday
rotation to Faraday oscillation just follows from accepting the
physically reasonable idea that, in the same way as the eigenstates
for photons coupled to
$\sigma_+$ excitons have a $(\sigma_+,\sigma_-)$ polarization,
the ones coupled to $\v x$ excitons have a $(\v x,\v y)$ polarization,
this change being rather obvious from the Shiva diagrams [16]
which represent the interactions of a pump exciton and a virtual exciton
coupled to the probe.

We wish to thank C. Benoit \`{a} la Guillaume, M. Chamarro, C. Testelin
for valuable discussions, and M. Loday for her help.

\end{document}